\RequirePackage[2020-02-02]{latexrelease}
\documentclass[aps,reprint,amsmath,amssymb,pre,onecolumn,12pt]{revtex4-2}
\usepackage{graphicx}% Include figure files
\usepackage{dcolumn}% Align table columns on decimal point
\usepackage{hyperref}
	\usepackage{mathtools}
	\usepackage{amsmath}	
	\usepackage{graphicx,wrapfig,lipsum}
\hypersetup{
	colorlinks   = false,
	linkcolor=red,          % color of internal links (change box color with linkbordercolor)
	citecolor=green,        % color of links to bibliography
	filecolor=magenta,      % color of file links
	urlcolor=cyan 
}
\usepackage{geometry}
\usepackage[dvipsnames]{xcolor}
 
\usepackage{xr}
\usepackage{natbib}
\begin{document}
	\title{Alterations in electroosmotic slip velocity: combined effect of viscoelasticity and surface potential undulation}
	\author{Bimalendu Mahapatra}
	\author{Aditya Bandopadhyay}%
	\email{aditya@mech.iitkgp.ac.in}
	\affiliation{Department of Mechanical Engineering, Indian Institute of Technology Kharagpur, Kharagpur - 721302, West Bengal, India}%
	%\date{\today}
	
	\begin{abstract}
In computational models of microchannel flows, the Helmholtz-Smoluchowski slip velocity boundary condition is often used because it approximates the motion of the electric double layer without resolving the charge density profiles close to the walls while drastically reducing the computational effort needed for the flow model to be solved. Despite working well for straight channel flow of Newtonian fluids, the approximation does not work well for flow involving complex fluids and spatially varying surface potential distribution. To treat these effects using the slip velocity boundary condition, it is necessary to understand how the surface potential and fluid properties affect the slip velocity. The present analysis shows the existence of a modified electroosmotic slip velocity for viscoelastic fluids, which is strongly dependent upon Deborah number and viscosity ratio, and this modification differs significantly from the slip velocity of Newtonian fluids. An augmentation of fluid elasticity results in an asymmetric distribution of slip velocity. Nonintuitively, the modulation wavelength of the imposed surface potential contributes to changing the slip velocity magnitude and adding periodicity to the solution. The proposed electroosmotic slip velocity for viscoelastic fluid can be used in computational models of microchannel flows to approximate the motion of the electric double layer without resolving the charge density profiles close to the walls.
	\end{abstract}
	\maketitle
	\section{Introduction}\label{sec1}
	A trend toward miniaturizing systems has recently enabled the development of micro and nanofluidic systems that have found application in medical, pharmaceutical, and environmental applications \cite{shoji1990micropump,stone2004engineering}. In biomedical and pharmaceutical applications, such systems are immensely advantageous since they result in lower reagent consumption, less analysis time, and a higher degree of automation due to the small volume of fluids they handle \cite{lee2004micro,kaminski2017controlled}. The fluids handled by microfluidic devices are typically of diverse properties; they range from the simple Newtonian to the complex viscoelastic fluids. An interesting feature of many biofluids that are transported by such devices is that they exhibit viscoelastic behaviour. For instance, blood, saliva, synovial fluid, protein solutions, DNA solutions, etc. , have long chain molecular structures \cite{moyers2008non,fam2007rheological,oldroyd1958non,phan1978nonlinear}. To effectively manipulate the underlying transport processes, it is essential to possess a thorough understanding of the flow process associated with each variety of fluid.
	
	These miniaturized transports have easily incorporated the electroosmosis technique owing to the ease of integration, portability, noise-free operation, and lack of mechanically moving components compared to the classical pressure-driven transport \cite{wu2017fluid,jiang2017transient,mehboudi2014simulation}. The electroosmosis phenomenon is an electrokinetic transport phenomenon that occurs when charged liquids move relative to a charged substrate under external electrical fields \cite{masliyah2006electrokinetic,hunter2013zeta}. In such a transport process, an electrical double layer (EDL) forms in the interfacial region formed by a physicochemical reaction between the ionized liquid and the charged substrate. A strong electromotive force is exerted on the mobilized ions inside the ionic liquid by the externally applied electric field, which causes the ionic liquid to move outward, thereby producing the electroosmotic flow (EOF).
	
	There has been an abundance of theoretical and experimental investigations of the EOF in the literature for both Newtonian and non-Newtonian fluids over the past few decades \cite{bhattacharyya2019enhanced,song2018electrokinetic,hoshyargar2018hydrodynamic,thanjavur2015electric}. However, viscoelastic fluid models are the most appropriate for describing the electroosmotic flow of bio/polymeric fluids \cite{zhao2013electrokinetics,lv2021induced,thien1977new,mahapatra2022effect}. Surface charges at channel walls usually determine the nature of electroosmotic flows \cite{jimenez2019combined,yazdi2015steric,mahapatra2021effect,mahapatra2021numerical}. In addition to the surface charge, an electric potential at equilibrium, known as the zeta potential, can also be quantified. Surface potentials can be acquired by embedding electrodes to the channel wall apart from surface potential due to physicochemical interactions \cite{ramos1998ac}. Variable surface potentials could trigger interesting flow patterns because they would account for the quasi-linear terms of the constitutive equation \cite{ghosh2015electroosmosis,mahapatra2020electroosmosis,mahapatra2021microconfined}.
	
	It has been reported that the continuum methodology has been used to directly model the charge distributions in EDLs \cite{tang2003modeling,tang2004joule,horiuchi2004joule,mahapatra2021numerical}. Given the substantial difference between the channel size and the EDL width for aqueous flows, the EDL is only 10 nm thick, while microchannels can be up to 100 $\mu$m in width, the computational requirements are extensive for complex fluid flows through micro-channels. A compromise is often made by using the layer model\cite{macinnes2003prediction,zimmerman2006rheometry} to approximate fluid motion within an EDL that does not directly resolve the distribution of charges on walls or velocity close to them. Instead of a wall boundary condition, slip velocity boundaries are assumed, representing fluid velocity at the edge of the EDL. The slip boundary condition requires only the applied electric field to be calculated and the fluid motion to be simulated at the channel length scale, thus reducing the computational requirements.
	
	From the above discussion, it is reasonable to deduce that the interactions between fluid elasticity and surface charge heterogeneity are likely to be complex in the micro-confined EOF of a viscoelastic fluid. This complex interaction between physicochemical properties and fluid rheology can lead to significant alterations in electroosmotic slip velocity. Previously, \citet{ghosh2015electroosmosis} derived approximate analytical solutions of electroosmotic slip velocity solutions for transport of the Upper Convected Maxwell (UCM) fluid over charge modulated surfaces. However, their study is focused on the effect of relaxation time only and neglected the effect other time scales or viscosity ratio present in the viscoelastic fluid models. Later, \citet{mahapatra2020electroosmosis} obtained slip velocity solutions for Oldroyd-B fluid in the presence of charge modulated surfaces, however neglected the effect of higher order terms in the asymptotic series expansion. For a complete understanding of electroosmotic slip velocity modification, it is important to analyze both viscoelasticty as well as surface potential undulation considering higher order corrections in the asymptotic analysis.
	
	In the light of the above motivation, this work studies the flow of a viscoelastic fluid (specifically, the flow of an Oldroyd-B fluid) with a modulated surface potential using an analytical method. Considering the Debye-H\"uckel linearization, we use a double perturbation technique to obtain the asymptotic solution for slip velocity. Our focus has been on the ``thin EDL" limits since these represent a more practical and physically realistic paradigm in problems involving electroosmotic flows. In the present analysis, we mainly focus on highlighting the effects of fluid elasticity and physicochemical alterations on the electroosmotic slip velocity by considering the higher-order corrections of the asymptotic analysis.
	
	The present analysis reveals that a modified slip velocity for viscoelastic fluids can be obtained, which is strongly dependent upon Deborah number and viscosity/retardation ratio. Moreover, we find significant deviations in the slip velocity for the Oldroyd-B fluid when compared to a Newtonian fluid, especially in thin EDLs. Axial asymmetry in the slip velocity distribution becomes more evident as the Deborah number increases and the viscosity ratio decreases. The modulation wavelength of the imposed surface potential nonintuitively contributes significantly towards altering the slip velocity magnitude and bringing in periodicity to the solution for higher-order terms in the asymptotic series expansion.
	
	\section{Problem formulation}
	\begin{figure}[ht]
		\centering
		\includegraphics[scale=1.2]{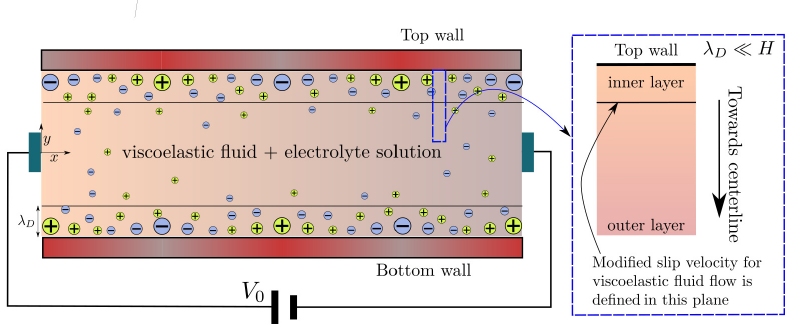}
		\caption{(Color online) Schematic highlighting the present physical system for the electroosmotic flow of viscoelastic fluid where color gradient at the channel wall depicts the variation of surface potential}
		\label{Fig:1}
	\end{figure}
	The present problem considers an electroosmotic flow of an Oldroyd-B fluid through a microchannel whose top and bottom walls have a separation distance of $2H$. The origin of the Cartesian coordinate system is situated at the channel centre-line, as shown in Fig.\ref{Fig:1}. Both the channel walls are physicochemically modulated and bear spatial variation of the surface potential of the form $\psi(\pm H)=\zeta_0(1+m \cos(n\frac{2\pi x}{d}+\theta))$. The wavelength of surface potential modulation is denoted as $n$, the magnitude of the surface potential as $m$, and the phase angle as $\theta$. The channel is filled with a quasi-linear viscoelastic fluid denoted by the Oldroyd-B constitutive model. An electrical double layer over the channel walls is formed due to the presence of a $z$:$z$ symmetric electrolyte in the viscoelastic fluid. The fluid is driven by an external electric field $E_0$ generated due to the imposition of potential difference $V_0$ between the inlet and outlet of the microchannel. We have obtained the slip velocity solutions by using the regular perturbation method in the inner layer. The higher-order correction terms is expected to impart a better insight into understanding the slip velocity variations due to the modifications in fluid or surface properties.

	A brief description of the parameters used in the analysis is necessary before going into the details of the issue. There are two orders of magnitude that determine the length ($L$) and height ($H$) of a microchannel: $L \sim H \sim O(10$ $\mu$m). As such, the relaxation and retardation time is of the order of $\lambda \sim \lambda_r \sim O(10^{-3}-0.2$ s) for the viscoelastic fluid. The bulk electrolyte concentration is $c_0 \sim 1$ mM, the absolute temperature is $T \sim 300$ K, the permittivity is $\epsilon \sim 7\times10^{-10}$ F/m, and the ionic diffusivity is $D \sim 2\times 10^{-9}$ m$^2$/s. A useful indicator of the viscoelastic fluid is the ratio $\frac{\lambda u_c}{H}$, referred to as the Deborah number ($De$), where $u_c$ indicates the velocity scale and $H$ indicates the length scale (here, it's considered to be the half channel height).
	\section{Asymptotic solution procedure}\label{sec:Asym_sol}
	We attempt to obtain the electroosmotic slip velocity solution by employing the asymptotic expansion of variables. We have carried out an order of magnitude analysis to obtain the magnitude of parametric values which will aid in taking assumptions for analytical simplicity. The characteristic velocity for the present analysis is considered to be same as the Smoluchowski velocity i.e. $u_c \sim u_{HS}$ which generally is of the order $O(10^{-4}-10^{-5})$ m/s. From the formulation of $De$ after substituting the parametric values, we obtain $De \sim O(10^{-3}-0.2)$. The thickness of electrical double layer which is also known as Debye length `$\lambda_D$' is of the order ($\lambda_D \sim 10$ nm). We define the ratio of Debye length to the channel half height as `$\delta$' which becomes $\delta=\lambda_D/H \sim 10^{-3}$. The ionic P\'eclet number defined by $Pe=u_c H/D$ becomes smaller than unity ($Pe<1$) which makes Poisson-Boltzmann equation valid to access the distribution of charge inside the microchannel. The surface potential considered in the present analysis is less than the thermal voltage which makes the Debye-H\"uckel linearization valid. The Reynolds number for the present system becomes of the order $Re \sim O(10^{-3})$ which leads us to neglect the inertial terms in the momentum equation as compared to the viscous terms. With the above mentioned simplifications, we present the governing equations for the present physical system. To distribution of the electrostatic potential will be governed by the linearized Poisson-Boltzmann equation given by
	\begin{equation}\label{eq:1}
		\small
		\frac{\partial ^{2}\psi }{\partial x ^{2}}+\frac{\partial ^{2}\psi }{\partial y ^{2}}=\frac{\psi }{\lambda_{D} ^{2}} \quad \text{where} \quad \lambda_D=\sqrt{\left( \frac{\epsilon k_B T}{2c_0z^2e^2} \right)}
	\end{equation}
	The boundary condition for the above equation is 
	\begin{equation}\label{eq:2}
		\small
		\psi(H)=\zeta_0\left(1+m_t \cos\left(n_t\frac{2\pi x}{d}+\theta_t\right)\right); ~~ \psi(-H)=\zeta_0\left(1+m_b \cos\left(n_b\frac{2\pi x}{d}+\theta_b\right)\right)
	\end{equation}
	The continuity equation and the momentum equation with the extra electrical body force term can be expressed as     \begin{equation}\label{eq:3}
		\nabla \cdot \mathbf{v}=0
	\end{equation}
	\begin{equation}\label{eq:4}
		-\mathbf{\nabla} p+\nabla \cdot \mathbf{\overline{\overline{\mathbf{\tau}}}}+\epsilon \nabla^2\psi (\nabla\psi-E_0)\hat{\mathbf{e_x}}=0
	\end{equation}
	It is essential to mention here that for the present system $\nabla\psi << E_0$ in the axial direction, the term $\nabla\psi$ can be neglected from the last term of Eq.\ref{eq:4}.
	The complete constitutive equation for the Oldroyd-B model is of the form
	\begin{equation}\label{eq:5}
		\overline{\overline{\mathbf{\tau}}}+\lambda\overline{\overline\tau}^\nabla = 2\eta\left(\overline{\overline{\textbf{D}}}+\lambda_r\overline{\overline{\textbf{D}}}^\nabla\right)
	\end{equation}
	Here $E_0$ denotes the axial external electric field. Fluid relaxation time and viscosity are denoted with $\lambda$ and $\eta$, respectively. In the current study, $\eta=\eta_s+\eta_p$, where $\eta_s$ and $\eta_p$ are the solvent and polymeric viscosity, respectively. The retardation time is $\lambda_r$ and $\beta=\eta_s/\eta=\lambda_r/\lambda$ is the viscosity ratio/retardation ratio. The deformation rate tensor is \textbf{D} and the upper convected derivative is $\overline{\overline{\textbf{A}}}^\nabla$, which is defined as $\overline{\overline{\textbf{A}}}^\nabla$=$\frac{D\overline{\overline{\textbf{A}}}}{Dt}$-$\overline{\overline{\textbf{A}}}\cdot\mathbf{\nabla}\textbf{u}$-$\mathbf{\nabla}\textbf{u}^T\cdot\overline{\overline{\textbf{A}}}$, where $\overline{\overline{\textbf{A}}}$ is a second order tensor. At the walls, the velocity field meets the no-slip boundary condition and no-penetration boundary condition, namely $\mathbf{v} = 0$ at $y = \pm H$. Physical parameter values indicate that $\lambda_D \ll H$, which results in the parameter $\delta \ll 1$. Applied to simple electroosmotic unidirectional flows, this achieves the thin EDL limit where the velocity profile changes rapidly from zero at the wall to the bulk velocity within a very short distance of $\sim O(\lambda_D)$. Following the above discussion, it can be inferred that flow domain consists of two distinct layers, as shown in Fig. \ref{Fig:1}. The one that is near the wall has a length scale $\sim O(\lambda_D)$ (``inner layer"), and the other is the bulk fluid outside of the EDL with nearly zero net charges (``outer layer"). A driving force is produced in the inner layer by the electric field. Motion in the outer layer, however, is determined by the viscous stress applied to it because of motion in the inner layer.
	
	First, we relocate the coordinate system such that $y'=H+y$ so that the origin appears in the bottom wall of the $y'$ coordinate before we can find the asymptotic solution for the inner layer. Then we proceed to obtain the dimensionless form of the relevant governing equations by employing the below mentioned non-dimensional scheme, in the inner layer where $\widetilde \Upsilon$ is the inner layer variable of a generic variable $\Upsilon$ and $\Upsilon_c$ is the characteristic variable used for non-dimensionalization.
	\begin{equation*}
		\small
		\begin{aligned}
			x_c = \frac{d}{2\pi};\quad y'_c = \lambda_D=\delta x_c;\quad %Y=\frac{y'}{x_c}\quad 
			u_c = u_{HS};\quad v_c = \delta u_{HS};\quad p_c = \frac{\eta u_{HS}}{x_c}\\ 
			\tau_{xy,c} = \frac{\eta u_{HS}}{\lambda_D};\quad \tau_{xx,c} = \frac{\tau_{xy,c}}{\delta} = \frac{\eta u_{HS}}{\delta  \lambda_D};\quad \tau_{yy,c} = \delta \tau_{xy,c} = \frac{\delta \eta u_{HS}}{\lambda_D};\quad  De=\frac{\lambda u_{HS}}{x_c}
		\end{aligned}
	\end{equation*}
	
	The resulting dimensionless equations of  only leading order terms in $\delta$ as $\delta \ll 1$ after the imposition of non-dimensional scheme to Eq.\ref{eq:1} and \ref{eq:3}-\ref{eq:5}, becomes 
	\begin{equation}\label{eq:21}
		\small
		\frac{\partial^2 \widetilde \psi}{\partial \widetilde y^2}=\widetilde \psi
	\end{equation}
	\begin{equation}\label{eq:22}
		\small
		\frac{\partial \widetilde \tau_{xx}}{\partial \widetilde x}+\frac{\partial \widetilde \tau_{xy}}{\partial \widetilde y}-\frac{\partial^2 \widetilde \psi}{\partial \widetilde y^2}=0
	\end{equation}
	\begin{equation}\label{eq:23}
		\small
		-\frac{\partial \widetilde p}{\partial \widetilde y}+\frac{\partial \widetilde \tau_{xy}}{\partial \widetilde x}+\frac{\partial \widetilde \tau_{yy}}{\partial \widetilde y}=0
	\end{equation}
	\begin{equation}\label{eq:24}
		\small
		\frac{\partial \widetilde u}{\partial \widetilde x}+\frac{\partial \widetilde v}{\partial \widetilde y}=0
	\end{equation}
	\begin{equation}\label{eq:25}
		\small
		\begin{aligned}
			\widetilde \tau_{xx}+De\bigg(\widetilde u\frac{\partial \widetilde \tau_{xx}}{\partial \widetilde x}+\widetilde v\frac{\partial \widetilde \tau_{xx}}{\partial \widetilde y}-2\bigg(\widetilde \tau_{xx}\frac{\partial \widetilde u}{\partial \widetilde x}+\widetilde \tau_{xy}\frac{\partial \widetilde u}{\partial \widetilde y}\bigg)\bigg)=-2De\beta\bigg(\frac{\partial \widetilde u}{\partial \widetilde y}\bigg)^2 
		\end{aligned}
	\end{equation}
	
	\begin{equation}\label{eq:26}
		\small
		\begin{aligned}
			\widetilde \tau_{xy}+De\bigg(\widetilde u\frac{\partial \widetilde \tau_{xy}}{\partial \widetilde x}+\widetilde v\frac{\partial \widetilde \tau_{xy}}{\partial \widetilde y}-\bigg(\widetilde \tau_{xx}\frac{\partial \widetilde v}{\partial \widetilde x}+\widetilde \tau_{yy}\frac{\partial \widetilde u}{\partial \widetilde y}\bigg)\bigg)=2\bigg[\frac{1}{2}\frac{\partial \widetilde u}{\partial \widetilde y}+De\beta\bigg(\frac{\widetilde u}{2}\frac{\partial^2 \widetilde u}{\partial \widetilde x\partial \widetilde y}\\+\frac{\widetilde v}{2}\frac{\partial^2 \widetilde u}{\partial \widetilde y^2}-\frac{\partial \widetilde u}{\partial \widetilde y}\frac{\partial \widetilde v}{\partial \widetilde y}\bigg)  \bigg]
		\end{aligned}
	\end{equation}
	
	\begin{equation}\label{eq:27}
		\small
		\begin{aligned}
			\widetilde \tau_{yy}+De\bigg(\widetilde u\frac{\partial \widetilde \tau_{yy}}{\partial \widetilde x}+\widetilde v\frac{\partial \widetilde \tau_{yy}}{\partial \widetilde y}-2\bigg(\widetilde \tau_{xy}\frac{\partial \widetilde v}{\partial \widetilde x}+\widetilde \tau_{yy}\frac{\partial \widetilde v}{\partial \widetilde y}\bigg)\bigg)=2\bigg[\frac{\partial \widetilde v}{\partial \widetilde y}+De\beta\bigg(\widetilde v\frac{\partial^2 \widetilde v}{\partial \widetilde y^2}+\widetilde u\frac{\partial^2 \widetilde v}{\partial \widetilde x \partial \widetilde y}\\-2\bigg(\bigg(\frac{\partial \widetilde v}{\partial \widetilde y}\bigg)^2+\frac{1}{2}\frac{\partial \widetilde u}{\partial \widetilde y} \cdot \frac{\partial \widetilde v}{\partial \widetilde x}  \bigg)\bigg)  \bigg]
		\end{aligned}
	\end{equation}
	the boundary conditions for velocity and potential at the wall is given by
	\begin{equation}
		\widetilde u(\widetilde y=0)=\widetilde v(\widetilde y=0)=0 \qquad  \text{and} \qquad  \widetilde \psi(\widetilde y=0)=1+m\cos(n\widetilde x+\theta)
	\end{equation}
	
	The exact solution to the aforementioned governing equations (Eq.\ref{eq:21}-\ref{eq:27}) in the inner layer is not feasible. Thus, we proceed to obtain an asymptotic solution using regular perturbation technique where $De$ and $\beta$ are considered as the perturbation parameters. For any variable $\widetilde\Upsilon$ in the inner layer, we expand the variable as
	\begin{equation}
		\begin{aligned}
			\widetilde\Upsilon=\widetilde\Upsilon_0+De\widetilde\Upsilon_{10}+\beta\widetilde\Upsilon_{01}+De\beta\widetilde\Upsilon_{11}+De^2\widetilde\Upsilon_{20}+\beta^2\widetilde\Upsilon_{02}+De^2\beta\widetilde\Upsilon_{21}\\+De\beta^2\widetilde\Upsilon_{12}+De^2\beta^2\widetilde\Upsilon_{22}+\cdot\cdot\cdot
		\end{aligned}
	\end{equation}
	
	\paragraph{Solution for potential distribution}
	It is important to note that the asymptotic series expansion for potential `$\widetilde \psi$' is not applicable since the potential distribution is independent of the fluid property for $Pe<1$. Thus, $\widetilde \psi$ is not dependent on $De$ and $\beta$ which represent the fluid properties and used as gauge function for series expansion. We directly obtain the solution for potential in the inner layer by solving the Eq.\ref{eq:21} which becomes
	\begin{equation}\label{eq:15}
		\widetilde{\psi}=[1+m\cos(n\widetilde{x}+\theta)] e^{-\widetilde y}
	\end{equation}      
	\paragraph{Solution for Leading order or $O(1)$ velocity}\label{sec:Sol_leadingorder} To obtain the leading order solution for velocity, first we access the leading order stress components from Eq.\ref{eq:25}-\ref{eq:27} which are of the form
	\begin{equation}\label{eq:45}
		\widetilde{\tau}_{xx,0}=0, \widetilde{\tau}_{xy,0}=\frac{\partial \widetilde u_0}{\partial \widetilde y}, \widetilde{\tau}_{yy,0}=2\frac{\partial \widetilde v_0}{\partial \widetilde y}
	\end{equation}
	we solve the axial momentum equation (Eq.\ref{eq:22}) by substituting the solution for potential (Eq.\ref{eq:15}) and the leading order stress components (Eq.\ref{eq:45}), which gives the expression for axial velocity in the inner layer
	\begin{equation}\label{eq:46}
		\widetilde u_0=(1+m\cos(n\widetilde{x}+\theta)) e^{-\widetilde y}+c_1(\widetilde{x})\widetilde{y}+c_2(\widetilde{x})
	\end{equation}
	after imposing the no slip boundary condition at $\widetilde y=0$ to Eq.\ref{eq:46} and assuming the solution is bounded for $\widetilde y \gg 1$, we obtain  $c_1(\widetilde{x})=0$ and  $c_2(\widetilde{x})=-(1+m\cos(n\widetilde{x}+\theta))$. Using the obtained expression for $c_1$ and $c_2$, the complete solution for the inner layer velocity in the leading order becomes
	\begin{equation}\label{eq:47}
		\widetilde u_0=(1+m\cos(n\widetilde{x}+\theta)) (e^{-\widetilde y}-1)
	\end{equation}
	The slip velocity is defined as the axial velocity at the interface of inner and outer layer. Thus, the leading order slip velocity will be equal to the inner layer axial velocity at $\widetilde y \to \infty$. The solution of leading order slip velocity is of the form
	\begin{equation}\label{eq:49}
		\lim_{\widetilde y >>1} \widetilde{u}_0=\lim_{\overline{y} \to -\overline H} \overline{u}_0=-(1+m\cos(n\widetilde{x}+\theta))
	\end{equation}
	The transverse velocity in the inner layer can be obtained by integrating the continuity Eq.\ref{eq:24} and using no penetration boundary condition. It is worth mentioning that, for $\delta \ll 1$ the boundary condition for transverse velocity to solve the outer layer will be no penetration boundary condition (i.e. $\overline v_0=0$ at $\overline y=\pm H$ ) because the matching condition at the interface becomes $\lim_{\overline{y} \to -\overline H} \overline{v}_0=\lim_{\widetilde y >>1} \delta\widetilde{v}_0 \to 0$. Then we proceed to obtain various orders of corrections to the leading order velocity, which will take the viscoelastic behaviour of the fluid into account.
	\paragraph{$O(De)$, $O(\beta)$ and $O(De\beta)$ corrections}\label{sec:Sol_O(De)}  
	The solution steps for various order of correction will be similar to the leading order solution, hence we will not mention those here explicitly. We first assess the $O(De)$ stress components in the inner layer, which becomes
	\begin{equation}
		\begin{aligned}
			\widetilde\tau_{xx,10} =-\widetilde u_{0}{\frac {\partial\widetilde \tau_{xx,0} }{\partial \widetilde x}}-\widetilde v_{0}  {\frac {\partial\widetilde \tau_{xx,0} }{\partial \widetilde y}}  +2\,\widetilde \tau_{xx,0}  {\frac{\partial \widetilde u_{0}}{\partial \widetilde x}}  +2\,\widetilde \tau_{xy,0}{\frac {\partial \widetilde u_{0} }{\partial \widetilde y}}  \\
			\widetilde \tau_{xy,10}={\frac {\partial \widetilde u_{10}}{\partial \widetilde y}}-\widetilde u_{0}{\frac {\partial \widetilde \tau_{xy,0}}{\partial \widetilde x}}-\widetilde v_{0}{\frac {\partial\widetilde \tau_{xy,0} }{\partial \widetilde y}}+\widetilde \tau_{xx,0}{\frac {\partial \widetilde v_{0} }{\partial \widetilde x}}+\widetilde \tau_{yy,0}{\frac {\partial \widetilde u_{0}}{\partial \widetilde y}}  \\
			\widetilde \tau_{yy,10}=2\,{\frac {\partial \widetilde v_{10}}{\partial \widetilde y}}-\widetilde u_{0}{\frac {\partial \widetilde \tau_{yy,0}}{\partial \widetilde x}}-\widetilde v_{0} {\frac {\partial\widetilde \tau_{yy,0} }{\partial \widetilde y}}+2\,\widetilde \tau_{xy,0}{\frac {\partial \widetilde v_{0} }{\partial \widetilde x}}+2\,\widetilde \tau_{yy,0}{\frac {\partial \widetilde v_{0} }{\partial \widetilde y}} 
		\end{aligned}
	\end{equation}
	one can observe that $\widetilde\tau_{xx,10}$ is only dependent on the leading order stress and velocity components, whereas $\widetilde \tau_{xy,10}$ and $\widetilde \tau_{yy,10}$ have dependency on the leading order terms as well as the $O(De)$ velocity components. We have solved the axial momentum equation and substituted the expressions for stress components and relevant boundary conditions to obtain the $O(De)$ slip velocity of the form
	\begin{equation}\label{eq:90}
		\begin{aligned}
			\lim_{\widetilde y >>1}\widetilde u_{10}=-\frac{3}{2}m(m
			\sin ( 2\,n\widetilde x+2\,\theta) +2\,\sin ( n\widetilde x+\theta )) n\\
		\end{aligned}
	\end{equation}
	Towards obtating the $O(\beta)$ velocity corrections, we obtain the constitutive relation for the $O(\beta)$ stress components in the inner layer, which reads
	\begin{equation}
		\begin{aligned}
			\widetilde \tau_{xx,01}=0; \quad \widetilde \tau_{xy,01}=\frac{\partial \widetilde u_{01}}{\partial \widetilde y}; \quad \widetilde \tau_{yy,01}=2\frac{\partial \widetilde v_{01}}{\partial \widetilde y}
		\end{aligned}
	\end{equation}
	here, $\widetilde\tau_{xx,10}$ becomes zero and $\widetilde \tau_{xy,10}$ and $\widetilde \tau_{yy,10}$ have dependency on $O(\beta)$ velocity components only. Following the similar solution procedure as the leading order solution, we have obtained trivial solution of slip velocity coefficients i.e., $\lim_{\widetilde y >>1}\widetilde u_{01}=0$ for $O(\beta)$. The absence of slip velocity contribution of $O(\beta)$ term physically mean that the flow does not have a sole dependency on the viscosity ratio `$\beta$'. To check whether the viscosity ratio has any impact on the solution when combined with the Deborah number we proceed to assess the $O(De\beta)$ velocity corrections. Obtained constitutive relation for the $O(De\beta)$ stress components in the inner layer is of the form
	\begin{equation}
		\begin{aligned}
			\widetilde\tau_{xx,1}=-2\left(\frac{\partial \widetilde u_0}{\partial \widetilde y}\right)^2+2\frac{\partial \widetilde u_0}{\partial \widetilde x}\widetilde\tau_{xx,01}+2\frac{\partial \widetilde u_0}{\partial \widetilde y}\widetilde\tau_{xy,01}+2\frac{\partial \widetilde u_{01}}{\partial \widetilde x}\widetilde\tau_{xx,0}+2\frac{\partial \widetilde u_{01}}{\partial \widetilde y}\widetilde\tau_{xy,0}\\-\frac{\partial \widetilde\tau_{xx,0}}{\partial \widetilde x}\widetilde u_{01}-\frac{\partial \widetilde \tau_{xx,0}}{\partial \widetilde y}\widetilde v_{01}-\frac{\partial \widetilde \tau_{xx,01}}{\partial \widetilde x}\widetilde u_{0}-\frac{\partial \widetilde \tau_{xx,01}}{\partial \widetilde y}\widetilde v_{0};\\
			\widetilde\tau_{xy,1}=\frac{\partial \widetilde u_1}{\partial \widetilde y}-2\frac{\partial \widetilde u_0}{\partial \widetilde y}\frac{\partial \widetilde v_0}{\partial \widetilde y}+\frac{\partial^2 \widetilde u_0}{\partial \widetilde y \partial \widetilde x}\widetilde u_0+\frac{\partial^2 \widetilde u_0}{\partial^2 \widetilde y}\widetilde v_0+\frac{\partial \widetilde u_0}{\partial \widetilde y}\widetilde \tau_{yy,01}+\frac{\partial \widetilde u_{01}}{\partial \widetilde y}\widetilde \tau_{yy,0}\\+\frac{\partial \widetilde v_0}{\partial \widetilde x}\widetilde \tau_{xx,01}+\frac{\partial \widetilde v_{01}}{\partial \widetilde x}\widetilde \tau_{xx,0}-\frac{\partial \widetilde \tau_{xy,0}}{\partial \widetilde x}\widetilde u_{01}-\frac{\partial \widetilde \tau_{xy,0}}{\partial \widetilde y}\widetilde v_{01}-\frac{\partial \widetilde \tau_{xy,01}}{\partial \widetilde x}\widetilde u_{0}-\frac{\partial \widetilde \tau_{xy,01}}{\partial \widetilde y}\widetilde v_{0};\\
			\widetilde\tau_{yy,1}=2\frac{\partial \widetilde v_1}{\partial \widetilde y}+2\frac{\partial^2 \widetilde v_0}{\partial \widetilde y \partial \widetilde x}\widetilde u_0+2\frac{\partial^2 \widetilde v_0}{\partial^2 \widetilde y}\widetilde v_0-4\left(\frac{\partial \widetilde v_0}{\partial \widetilde y}\right)^2-2\frac{\partial \widetilde u_0}{\partial \widetilde y}\frac{\partial \widetilde v_0}{\partial \widetilde x}\\+2\frac{\partial \widetilde v_0}{\partial \widetilde x}\tau_{xy,01}+2\frac{\partial \widetilde v_0}{\partial \widetilde y}\widetilde \tau_{yy,01}+2\frac{\partial \widetilde v_{01}}{\partial \widetilde x}\widetilde \tau_{xy,0}+2\frac{\partial \widetilde v_{01}}{\partial \widetilde y}\widetilde \tau_{yy,0}-\frac{\partial \widetilde \tau_{yy,0}}{\partial \widetilde x}\widetilde u_{01}\\-\frac{\partial \widetilde \tau_{yy,0}}{\partial \widetilde y}\widetilde v_{01}-\frac{\partial \widetilde \tau_{yy,01}}{\partial \widetilde x}\widetilde u_{0}-\frac{\partial \widetilde \tau_{yy,01}}{\partial \widetilde y}\widetilde v_{0}
		\end{aligned}
	\end{equation} 
	the $O(De\beta)$ normal stress component i.e. $\widetilde\tau_{xx,1}$ depends on leading order and $O(\beta)$ velocity and stress components, whereas $\widetilde\tau_{xy,1}$ and $\widetilde\tau_{yy,1}$ have dependency on the $O(De\beta)$ velocity components along with the leading order and $O(\beta)$ velocity and stress components. Solving for velocity, the $O(De\beta)$ correction becomes
	\begin{equation}
		\begin{aligned}\label{eq:91}
			\lim_{\widetilde y >>1}\widetilde u_{11}=\frac{3}{2}m(
			m\sin (2\,n\widetilde x+2\,\theta) +2\,\sin ( n\widetilde x+\theta )) n\\
		\end{aligned}
	\end{equation} 
	Here it is worth mentioning that, $\beta$ is considered as a gauge function in the analysis and we have obtained a zero correction velocity for $O(\beta)$. While deriving the velocity corrections for $O(De\beta)$ terms, $\beta$ is considered asymptotically small. However, for comparison purpose if $\beta \to 1$, the validity of the present analysis will still hold good for $O(De\beta)$ and other higher order terms [$O(De^p\beta^q)$, where $p,q=2,3,4,....$] in the expansion series due to the fact that $De$ is asymptotically small and the maximum value of $\beta^q$ can be 1. 
	
	Interestingly, the slip velocity correction coefficients for $O(De)$ and $O(De\beta)$ are exactly equal and opposite. The solution for slip velocity will be exactly equal and opposite for $\beta=1$ considering only $O(De)$ and $O(De\beta)$ terms and excluding the leading order terms. This physically means that the rheological behaviour of the viscoelastic fluid model will be exactly equal to the Newtonian fluid when $\beta \to 1$, which is true since there will be the only contribution of solvent viscosity as the viscosity ratio becomes unity. Another important aspect of the solution is the appearance of the parameter `$n$' defining the wavelength of charge modulation as a multiplication factor in Eq.\ref{eq:90} and Eq.\ref{eq:91}, whereas in the leading order solution (Eq.\ref{eq:49}) it is absent. The modulation wavelength of the imposed surface potential contributes significantly to altering the slip velocity magnitude and bringing in periodicity to the solution for $O(De)$ and $O(De\beta)$ terms.
	\paragraph{$O(\beta^2)$ and $O(De\beta^2)$ corrections}\label{sec:Sol_O(beta)} 
	We have obtained trivial solution of slip velocity coefficients for $O(\beta^2)$, and $O(De\beta^2)$ terms following the similar procedure as the leading order solution. We have presented the expression for the stress components employed to solve the system of equations in the Appendix \ref{sec:ASol_O(beta2)} and \ref{sec:ASol_O(Debeta2)} for $O(\beta^2)$, and $O(De\beta^2)$ corrections.
	\begin{equation}
		\begin{aligned}
			\lim_{\widetilde y >>1}\widetilde u_{02}=0; \quad \lim_{\widetilde y >>1} \widetilde u_{12}=0
		\end{aligned}
	\end{equation}
	The absence of slip velocity contribution of $O(\beta)$ and $O(\beta^2)$ terms physically mean that the flow does not have a sole dependency on the retardation time as $\beta=\eta_s/\eta=\lambda_r/\lambda$. Moreover, the relaxation time is the main governing parameter for the viscoelastic fluid flow, as we obtain a non-zero solution for $O(De\beta)$ terms, which indicates that the viscosity/retardation ratio combined with the Deborah number can contribute towards the slip velocity correction. The $O(De\beta^2)$ stress coefficients are dependent on the solutions of $O(\beta)$ and $O(\beta^2)$ terms (refer to the Appendix \ref{sec:ASol_O(Debeta2)}) which becomes zero, thus we obtain trivial solution for slip velocity correction coefficient of $O(De\beta^2)$. 
	\paragraph{$O(De^2)$, $O(De^2\beta)$, and $O(De^2\beta^2)$ corrections}\label{sec:Sol_O(De2)}  
	Similarly, we proceed to obtain the higher order corrections to the slip velocity coefficients by using the expressions for various stress components in the Appendix \ref{sec:ASol_O(De2)}, \ref{sec:ASol_O(De2beta)}, and \ref{sec:ASol_O(De2beta2)} for {$O(De^2)$, $O(De^2\beta)$, and $O(De^2\beta^2)$ terms, respectively. The solution for slip velocity correction coefficient of $O(De^2)$ becomes
		\begin{equation}\label{eq:200}
			\small
			\begin{aligned}
				\lim_{\widetilde y >>1}\widetilde u_{20}=65n^2m\bigg(m^2\cos(\theta)^3\cos(n\widetilde x)^3-m^2\cos(n\widetilde x)^2\sin(n\widetilde x)\sin(\theta)\cos(\theta)^2\\-\frac{3}{4}m^2\cos(\theta)\cos(n\widetilde x)^3+\frac{1}{4} m^2\cos(n \widetilde x)^2 \sin(\theta) \sin(n \widetilde x)-\frac{3}{4}\cos(\theta)^3\cos(n\widetilde x)m^2\\+\frac{1}{4} \sin(\theta) \sin(n \widetilde x) m^2\cos(\theta)^2+\frac{43}{65}m\cos(n\widetilde x)^2\cos(\theta)^2-\frac{43}{65}\cos(\theta)\cos(n\widetilde x)\sin(\theta)\\m\sin(n\widetilde x)+\frac{151}{260} \cos(\theta) \cos(n \widetilde x) m^2-\frac{21}{260}\sin(\theta)\sin(n\widetilde x)m^2-\frac{43}{130}m\cos(n\widetilde x)^2\\-\frac{43}{130}m\cos(\theta)^2+\frac{21}{260} \cos(\theta) \cos(n \widetilde x)-\frac{21}{260} \sin(\theta) sin(n \widetilde x)+\frac{21}{130}m\bigg)
			\end{aligned}
		\end{equation}
		and the solution for slip velocity correction coefficient of $O(De^2\beta)$ is of the form
		\begin{equation}\label{eq:201}
			\small
			\begin{aligned}
				\lim_{\widetilde y >>1}\widetilde u_{21}=-4mn^2\bigg(\frac{171}{4}m^2\cos(\theta)^3\cos(n\widetilde x)^3-\frac{171}{4}m^2\cos(n\widetilde x)^2\sin(n\widetilde x)\sin(\theta)\cos(\theta)^2\\-\frac{513}{16}\cos(\theta)^3\cos(n\widetilde x)m^2+\frac{171}{16}\sin(\theta) \sin(n \widetilde x) m^2 \cos(\theta)^2-\frac{513}{16}m^2\cos(\theta)\cos(n\widetilde x)^3\\+\frac{171}{16} m^2 \cos(n \widetilde x)^2 \sin(\theta) \sin(n \widetilde x)+\frac{113}{4} m \cos(n \widetilde x)^2 \cos(\theta)^2-\frac{113}{4}\cos(\theta)\cos(n\widetilde x)\\m\sin(\theta)\sin(n\widetilde x)+\frac{397}{16} \cos(\theta) \cos(n \widetilde x) m^2-\frac{55}{16}\sin(\theta)\sin(n\widetilde x)m^2-\frac{113}{8}m\cos(\theta)^2\\-\frac{113}{8}m\cos(n\widetilde x)^2+\frac{55}{16} \cos(\theta) \cos(n \widetilde x)-\frac{55}{16} \sin(\theta) \sin(n \widetilde x)+\frac{55}{8} m\bigg)\\
			\end{aligned}
		\end{equation}
		then we obtain the slip velocity correction coefficient of $O(De^2\beta^2)$, which reads
		\begin{equation}\label{eq:202}
			\small
			\begin{aligned}
				\lim_{\widetilde y >>1}\widetilde u_{22}=4n^2m\bigg(\frac{67}{4}\cos(\theta)^3m^2\cos(n\widetilde x)^3-\frac{67}{4}m^2\cos(n\widetilde x)^2\sin(n\widetilde x)\sin(\theta)\cos(\theta)^2\\-\frac{201}{16}\cos(\theta)^3\cos(n\widetilde x)m^2+\frac{67}{16}\sin(\theta) \sin(n \widetilde x) m^2 \cos(\theta)^2-\frac{201}{16}\cos(\theta)m^2\cos(n\widetilde x)^3\\+\frac{67}{16}m^2\cos(n\widetilde x)^2\sin(\theta)\sin(n\widetilde x)+\frac{91}{8}m\cos(n\widetilde x)^2\cos(\theta)^2-\frac{91}{8}\cos(\theta)\cos(n\widetilde x)\\\sin(\theta)m\sin(n\widetilde x)+\frac{79}{8}\cos(\theta)\cos(n\widetilde x)m^2-\frac{3}{2}\sin(\theta)\sin(n\widetilde x)m^2-\frac{91}{16}m\cos(\theta)^2\\-\frac{91}{16}m\cos(n\widetilde x)^2+\frac{3}{2}\cos(\theta)\cos(n\widetilde x)-\frac{3}{2}\sin(\theta)\sin(n\widetilde x)+3m\bigg)
			\end{aligned}
		\end{equation}
		The above slip velocity correction terms involve the parameter `$n^2$' as multiplication factors, and we also observe that the solution consists of more periodic functions compared to the previously obtained lower-order correction terms. We will highlight the effect of individual correction terms on the overall solution and flow dynamics graphically in Sec. \ref{sec:Results}. The slip velocity solution, i.e., Eq. \ref{eq:125} from the asymptotic analysis, can be used as the moving wall velocity boundary condition at the channel walls.
		\begin{equation}\label{eq:125}
			\begin{aligned}
				\overline u_s=\lim_{\widetilde y >>1}(\widetilde u_0+De\widetilde u_{10}+\beta\widetilde u_{01}+De\beta\widetilde u_{11}+De^2\widetilde u_{20}+\beta^2\widetilde u_{02}+De^2\beta\widetilde u_{21}\\+De\beta^2\widetilde u_{12}+De^2\beta^2\widetilde u_{22})
			\end{aligned}
		\end{equation}
		\section{Results and discussion}\label{sec:Results}
		In the previous section (Sec. \ref{sec:Asym_sol}), we have discussed the procedure to obtain the slip velocity using a double perturbation methods for thin electrical double layer. This section highlights the electroosmotic slip velocity variations with changing fluid elasticity and physicochemical properties of the channel wall considering the higher-order corrections.
		\subsection{Validation of the Analytical solution}
		Before proceeding toward the results, it is essential to validate the analytical framework used to obtain the modified electroosmotic slip velocity for viscoelastic fluid flow under thin EDL consideration. We have compared the slip velocity solution obtained from our present asymptotic method with existing solutions from literature\cite{ghosh2015electroosmosis,mahapatra2020electroosmosis}. For varying Deborah number, the comparison plots for electroosmotic slip velocity are presented in Fig. \ref{Fig:2}. We found that our present asymptotic solution has an excellent agreement with existing slip velocity solutions considering the leading order and first order solutions in $De$ i.e., $\overline u_s=\lim_{\widetilde y >>1}(\widetilde u_0+De\widetilde u_{10})$. Generally, the electroomotic flow of viscoelastic fluid through the microchannel has $De \sim O(10^{-3}-0.4)$\cite{ghosh2015electroosmosis}, which our analytical solution is able to predict successfully. 
		\begin{figure}[ht]
			\centering
			\includegraphics[scale=0.4]{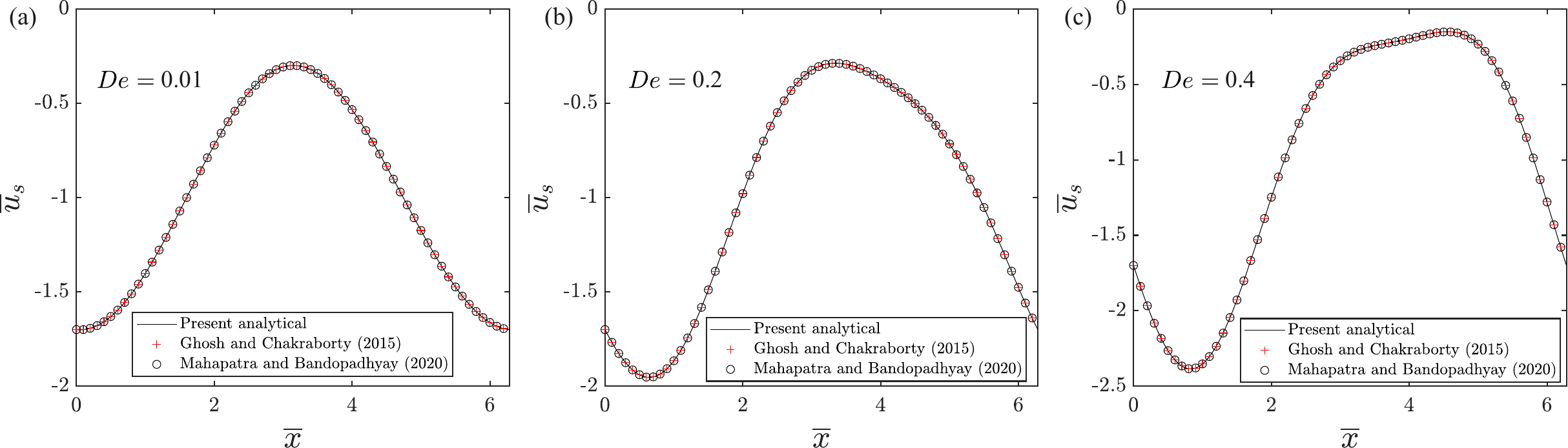}
			\caption{(Color online) Comparison of the slip velocity profiles ($\overline u_s=\lim_{\widetilde y >>1}(\widetilde u_0+De\widetilde u_{10}$) obtained from the present analytical solution considering a symmetric system with the asymptotic solution from literature\cite{ghosh2015electroosmosis,mahapatra2020electroosmosis} for varying Deborah number, (a) $De=0.01$, (b) $De=0.2$, (c) $De=0.4$}
			\label{Fig:2}
		\end{figure}
		We can observe that, for lower Deborah number, the electroosmotic slip velocity is symmetric about $\overline x=3.141$, however, as $De$ increases asymmetricity is introduced due to the inclusion of $O(De)$ non-linear terms. The present analytical solution largely depends on the fluid properties such as relaxation time and retardation time, which are non-dimensionally defined as Deborah number $De$ and retardation ratio $\beta$. It is worth mentioning that, for validation purpose we have considered the electroosmotic slip velocity to be $\overline u_s=\lim_{\widetilde y >>1}(\widetilde u_0+De\widetilde u_{10})$ due to the fact that the existing solution considered only the leading order and $O(De)$ terms \cite{mahapatra2020electroosmosis}. However, in the present analysis, we have considered up to nine terms in the asymptotic expansion series defined as $\overline u_s=\lim_{\widetilde y >>1}(\widetilde u_0+De\widetilde u_{10}+\beta\widetilde u_{01}+De\beta\widetilde u_{11}+De^2\widetilde u_{20}+\beta^2\widetilde u_{02}+De^2\beta\widetilde u_{21}+De\beta^2\widetilde u_{12}+De^2\beta^2\widetilde u_{22})$ while deriving the solution and also included the effect of all the correction terms for presenting important parametric variations.
		\subsection{Effect of viscoelasticity}
		\begin{figure}[ht]
			\centering
			\includegraphics[scale=0.6]{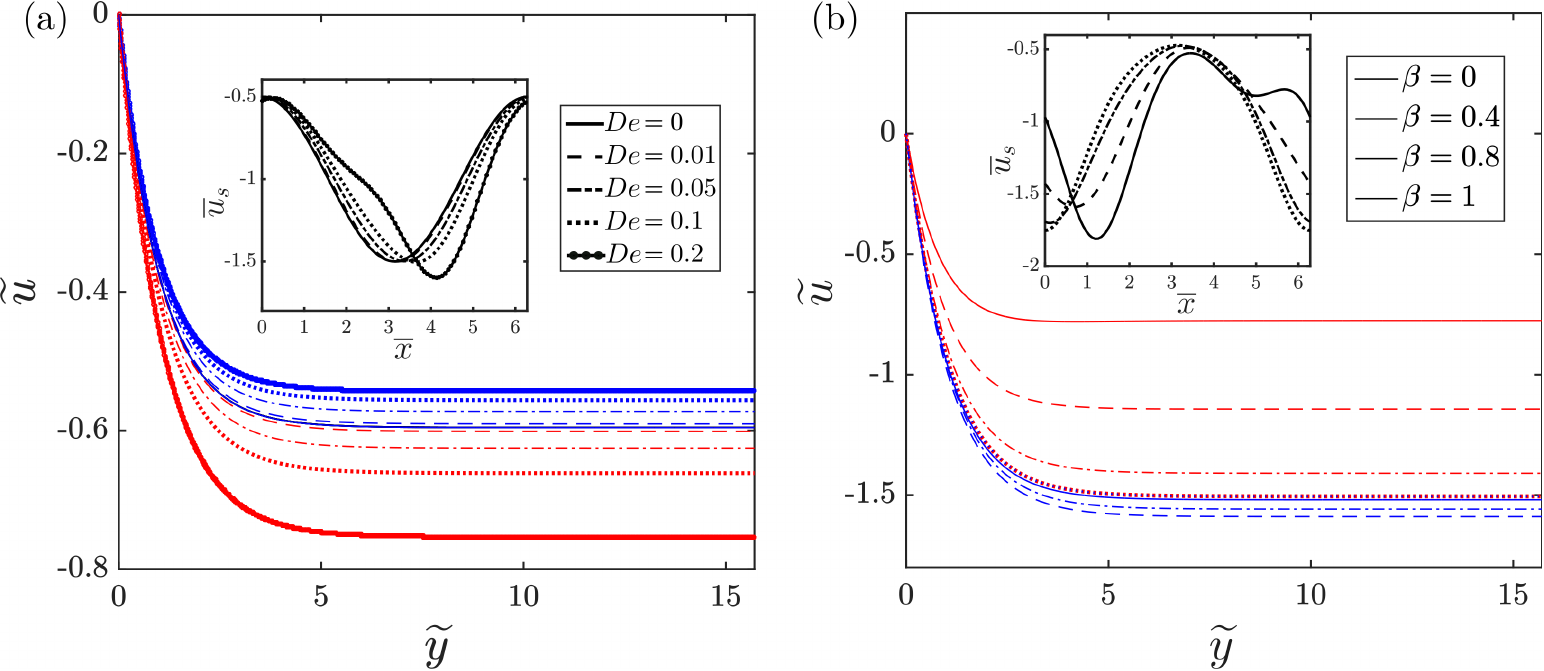}
			\caption{(Color online) (a) Variation of inner layer axial velocity $\widetilde u$ along the inner layer variable for transverse direction $\widetilde y$ at two axial positions $\overline x=\pi/5$ (denoted as solid lines) and $\overline x=9\pi/5$ (denoted as dashed lines), the inset shows the variation of slip velocity $\overline u_s$ vs $\overline x$ for varying $De$ at $\beta=0$, $m=0.5$, $n=1$ and $\theta=\pi$, (b) $\widetilde u$ vs $\widetilde y$ at $\overline x=\pi/5$ (denoted as solid lines) and $\overline x=9\pi/5$ (denoted as dashed lines), the inset highlights $\overline u_s$ vs $\overline x$ for varying $\beta$ at $De=0.3$, $m=0.5$, $n=1$ and $\theta=0$}
			\label{Fig:3}
		\end{figure}
		
		In viscoelastic fluid flows, changes in fluid rheology can greatly alter the flow structures, and for fluid transport involving electric fields, the alterations in electroosmotic slip velocity can be influenced remarkably. In this light, we have attempted to illustrate the effect of fluid rheology, non-dimensionally denoted by Deborah number `$De$' and viscosity/retardation ratio `$\beta$' on the variation of inner layer axial velocity and electroosmotic slip velocity. We have observed that (refer to inset of Fig. \ref{Fig:3}(a)) with increase in fluid elasticity or $De$, the slip velocity distribution which was symmetric about $\overline x=3.141$ for Newtonian fluid breaks. This symmetry breaking phenomenon is also reflected in the inner layer axial velocity distribution plotted at two equidistant points about $\overline x=3.141$ (symmetry line) in Fig. \ref{Fig:3}(a). Augmentation in the electroosmotic slip velocity $\lvert\overline u_s\lvert$ is also observed as a result of increased viscoelasticity. The viscosity/retardation ratio also plays an crucial role towards determining the effective viscoelasticty of the complex fluid. As we increase the viscosity ratio $\beta$ keeping the Deborah number constant, a reduced periodicity is obtained in the slip velocity distribution as shown in the inset of Fig. \ref{Fig:3}(b). From Fig. \ref{Fig:3}(b), we also observe that the slip velocity distribution which was asymmetric about $\overline x=3.141$ for lower $\beta$, becomes symmetric with increase in $\beta$. This occurs due to the fact that a higher viscosity ratio increases the solvent contribution in the viscoelastic fluid, causing it to behave like Newtonian fluid. A viscoelastic fluid behaves as a purely Newtonian fluid when $\beta = 1$ which is the maximum value $\beta$ can attain, and for $\beta = 1$ the total viscosity equals the solvent viscosity. It can be noted that, the effect of relaxation time and retardation time in the viscoelastic fluid is physically opposite to each other. This behavior is reflected in the non-dimensional system where the effect of varying $De$ and $\beta$ on the slip velocity distribution is qualitatively opposite as discussed.

		\subsection{Effect of undulated surface potential }\label{subsec:EMW}
		\begin{figure}
			\begin{minipage}[c]{0.3\textwidth}
				\includegraphics[scale=0.62]{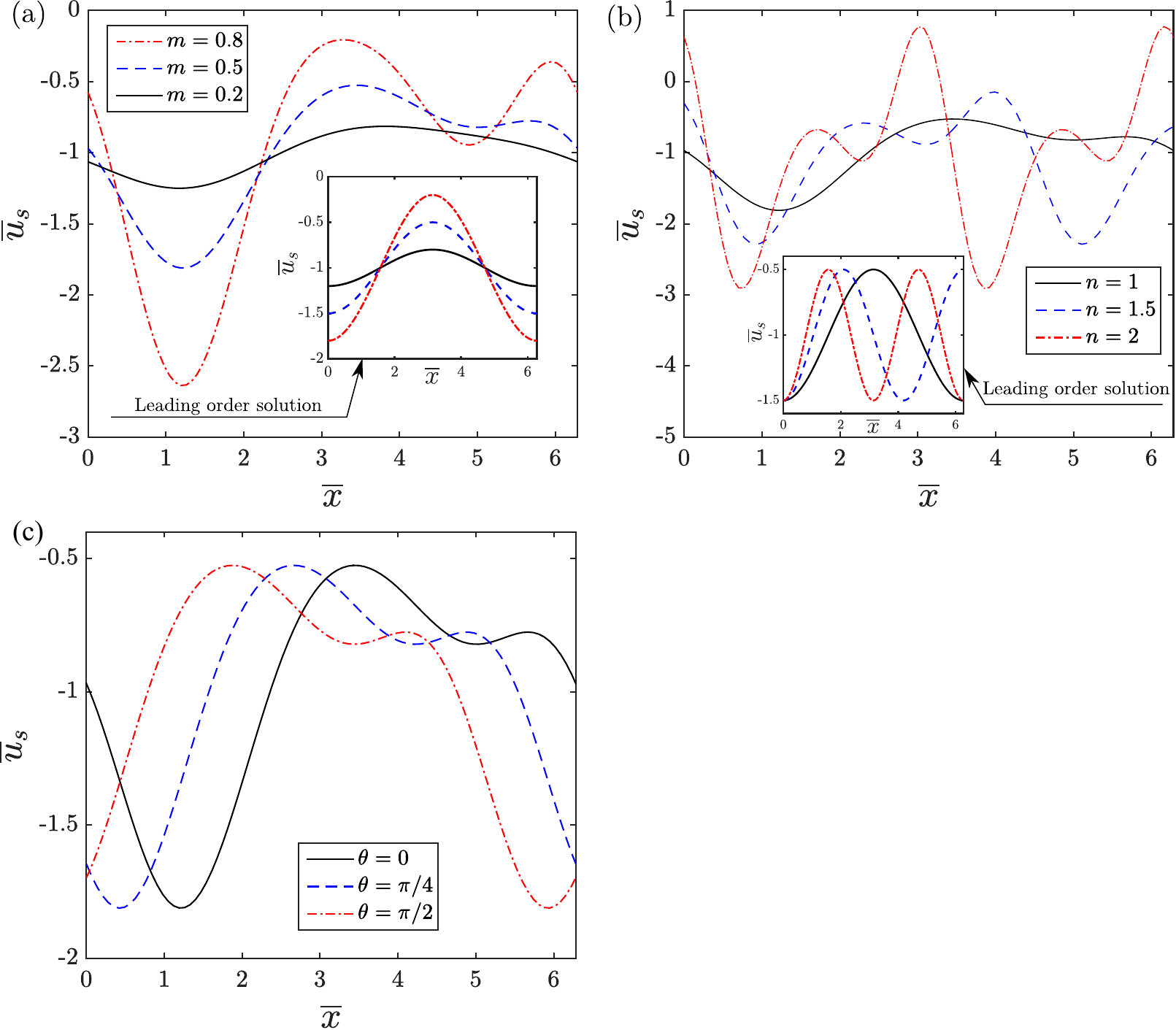}
			\end{minipage}\hfill
			\begin{minipage}[c]{0.45\textwidth}
				\begin{flushleft}
					\vspace{5cm}
					\caption{(Color online) (a) $\overline u_s$ vs $\overline x$ for varying $m$, the inset shows the slip velocity variation considering only leading order terms at $De=0.3$, $\beta=0$, $n=1$ and $\theta=0$, (b) $\overline u_s$ vs $\overline x$ for varying $n$, the inset shows the slip velocity variation considering only leading order terms at $De=0.3$, $\beta=0$, $m=0.5$ and $\theta=0$, (c) $\overline u_s$ vs $\overline x$ for asymmetric system at $De=0.3$, $\beta=0$, $m=0.5$ and $n=1$,}
					\label{Fig:4}
				\end{flushleft}
			\end{minipage}
		\end{figure}
		The physicochemical properties of fluid-solid interfaces, such as the surface charge distribution at the interface, significantly govern the electroosmotic flow of complex fluids. In the present study, surface charge variation is achieved by changing the magnitude, wavelength and phase angle of the imposed surface potential. First, we highlight the effect of changing strength of the surface potential denoted by `$m$' on the electroosmotic slip velocity variation. For a Newtonian fluid, it is intuitive to infer that with increasing strength of the surface potential, the strength of the slip velocity increases without affecting the periodicity. This intuitive observation is true for Newtonian fluid, as we have quantitatively shown that in the leading order (refer to inset of Fig. \ref{Fig:4} (a)) the magnitude of $\overline u_s$ only varies and the $\overline u_s$ vs $\overline x$ curve remains symmetric about the symmetry line ($\overline x=3.141$). The leading order solution gives the flow field of Newtonian fluid as all the non linear terms associated with $De$ and $\beta$ becomes zero. To understand this behavior better, let us consider the leading order solution for slip velocity i.e., $\lim_{\widetilde y >>1} \widetilde{u}_0=\lim_{\overline{y} \to -\overline H} \overline{u}_0=-(1+m\cos(n\widetilde{x}+\theta))$. Here one can observe that the parameter `$m$' is associated with only one periodic function. However, as the non-linearity increase with the inclusion of viscoelasticity in the fluid, there appears more periodic terms associated with `$m$' (refer to Eq.\ref{eq:90}, \ref{eq:91}, \ref{eq:200}-\ref{eq:202}). This non-intuitive behavior is more pronounced with increasing non-linearity in the flow which is a reflection of augmented viscoelasticity. It is worth mentioning that the increase in strength of surface potential also increases the magnitude of electroosmotic slip velocity along with bringing in periodic undulations only when the higher order corrections terms are taken into account.

		Although the imposed surface potential has only one period, i.e., $n=1$, we observed that there are more periodic undulations in the electroosmotic slip velocity with increasing surface potential strength. It leads to the analysis of how changing wavelengths of modulation impacts electroosmotic slip velocity. It is interesting to observe from Fig. \ref{Fig:4}(b) that the alterations in the parameter `$n$' employed to define the wavelength of surface potential distribution affect the slip velocity considerably. A decrease in wavelength, i.e., an increase in $n$, results in a more periodic slip velocity distribution and augments the slip velocity magnitude. This is also a non-intuitive feature of the present analysis that with changing modulation wavelength of surface potential we obtain magnitude modifications in electroosmotic slip velocity. However, for Newtonian fluids the scenario is inherent that the alterations in the value of `$n$' is directly reflected in the variation of $\overline u_s$ (refer to inset of Fig. \ref{Fig:4} (b)) without affecting the magnitude of $\overline u_s$. The augmentation in electroosmotic slip velocity with reducing wavelength can be analyzed mathematically considering various order of corrections. A conscious observation to the $O(De)$ slip velocity obtained as $\lim_{\widetilde y >>1}\widetilde u_{10}=-\frac{3}{2}mn(m
		\sin ( 2\,n\widetilde x+2\,\theta) +2\,\sin ( n\widetilde x+\theta ))$ revels that the parameter `$n$' is multiplied to the periodic terms and increasing this will consequently augment the slip velocity magnitude. The modulation wavelength $n$ exists as multiplication factor in $O(De)$ and $O(De\beta)$ terms, where as $n^2$ exists as multiplication factor in $O(De^2)$, $O(De^2\beta)$ and $O(De^2\beta^2)$ terms (refer to Eq.\ref{eq:90}, \ref{eq:91}, \ref{eq:200}-\ref{eq:202}). However, in the leading order solution the parameter $n$ was absent as a multiplication factor, which is the reason for zero slip velocity augmentation in Newtonian fluid. From the above discussion, it is clear that the higher-order corrections to the slip velocity non-intuitively bring in periodicity to the slip velocity distributions and are also responsible for the augmentation of slip velocity magnitudes. The above discussion dictates that the complex flow physics of the viscoelastic fluid flow can only be correctly assessed by considering the higher-order corrections, whereas the leading order solution only describes the Newtonian fluid flow features. 
		
		In an attempt to illustrate the effect of phase angle `$\theta$' on the slip velocity variation Fig. \ref{Fig:4}(c) is presented. From the figure, it is observed that the slip velocity profile is shifted by an equal phase angle as the imposed surface potential distribution without affecting the strength or the periodicity of the slip velocity distribution. Accordingly, the effect of phase angle is not accentuated in the higher-order corrections, unlike the alterations in magnitude and wavelength of the surface potential as previously discussed. 
		\subsection{Higher order dependency on electroosmotic slip velocity}
		\begin{figure}[ht]
			\centering
			\includegraphics[scale=0.4]{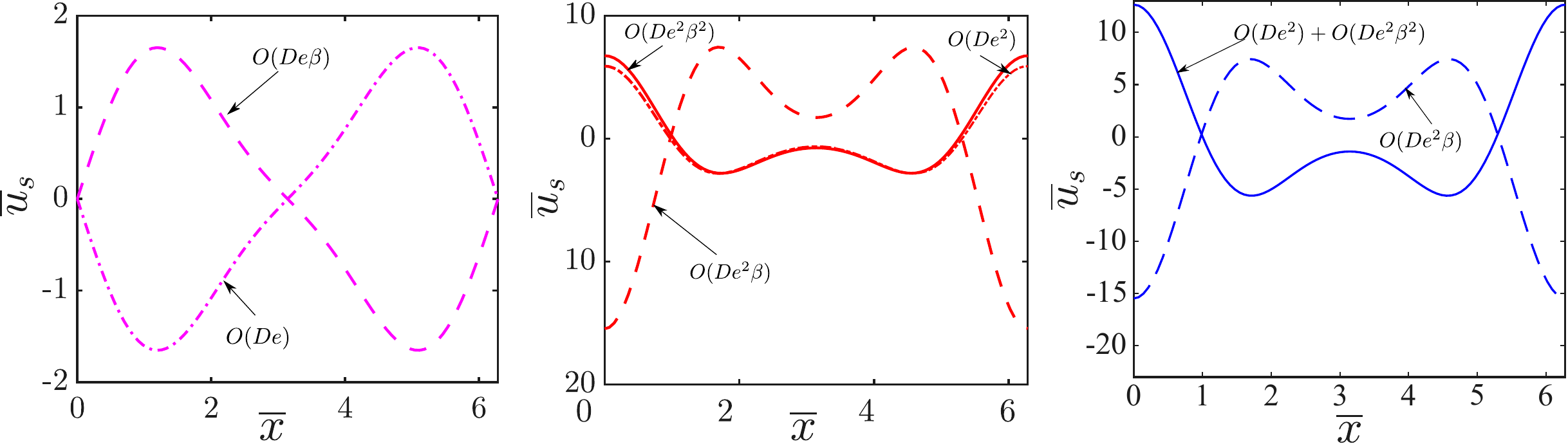}
			\caption{(Color online) Slip velocity corrections of various orders presented to understand the individual effect of parameters on the modified slip velocity}
			\label{Fig:5}
		\end{figure}
		The analytical solution obtained using the double perturbation analysis highlighted non-trivial higher-order corrections to the electroosmotic slip velocity as presented in Sec. \ref{sec:Asym_sol} and Sec. \ref{sec:Results}. This dictates that the higher-order corrections in an asymptotic analysis are crucial while analyzing the electroosmotic flow involving viscoelastic fluids. These higher-order correction terms tend to illustrate non-intuitive features which can not be accessed without proper analysis of individual correction terms. In Fig. \ref{Fig:5}, we have illustrated the correction coefficients of various terms at $\beta=1$ to have a better insight into individual contributions and relations with each other. When $\beta$ becomes unity, this physically means the complex fluid only has a solvent contribution, which rheologically behaves as a Newtonian fluid. Thus, we should only have the leading order solution for $\beta=1$, where the higher-order corrections do not have any contributions toward the complete slip velocity solution. We observe from the Fig. \ref{Fig:5} that the $O(De)$ and  $O(De\beta)$ correction coefficients are exactly equal and opposite (refer to Eq.\ref{eq:90}, \ref{eq:91}), and the summation of them becomes zero. The $O(De^2)$ and $O(De^2\beta^2)$ coefficients have similar axial variation, whereas $O(De^2\beta)$ coefficients have opposite axial variation with increased magnitude. The summation of $O(De^2)$ and $O(De^2\beta^2)$ terms is comparable but not exactly equal to $O(De^2\beta)$ terms (refer to refer to Eq. \ref{eq:200}-\ref{eq:202}). Thus, to have only the leading order solution at $\beta=1$, one needs to consider the contributions of all other higher-order terms in the asymptotic series expansion. From the above discussion, it appears that there will be small, but non-zero contributions from the other higher order terms in the slip velocity calculation. In practice, obtaining those higher order coefficients is not straightforward and analytically challenging. Consequently, the solution obtained considering the nine terms in the asymptotic series expansion can accurately be employed to predict the modified electroosmotic slip velocity for Oldroyd-B fluid flow in the presence of spatially varying surface potential.
		\section{Conclusion}
		To conclude, we have analytically derived the modified slip velocity solution at the vicinity of the channel wall for an electroosmotic flow of complex viscoelastic fluid described by the Oldroyd-B constitutive equation in the presence of surface charge modulation. The regular perturbation method was used to determine the electroosmotic slip velocity solution at the inner layer. We successfully validated the solutions of our asymptotic analysis against existing literature results. At the slip plane, we observe axial asymmetry of slip velocity distribution due to the enhanced fluid elasticity, which is a reflection of the elastic recoil of the fluid particles opposite to the flow direction. Significant alterations in the slip velocity magnitude are also reported for increasing viscoelasticity. A change in the strength and wavelength of the surface potential distribution is also responsible for a consequential change in the slip velocity. The electroosmotic slip velocity profile displays a phase shift without affecting its magnitude as a result of the imposed phase difference to the surface potential distribution. The non-intuitive features in the slip velocity variation for an electroosmotic flow of viscoelastic fluid are accessed by analyzing the individual higher-order correction terms. The present analysis proposes a modified electroosmotic slip velocity for a quasi-linear Oldroyd-B fluid, which can be used in computational models of microchannel flows to approximate the motion of the electric double layer without resolving the charge density profiles close to the walls. With the thin EDL approach, this will dramatically reduce the computational effort for electroosmotic flow model and viscoelastic fluid model solutions under varying surface charge conditions.

		%	\section*{Supporting Information}
		%	Supporting information is information that is not essential to the article, but provides greater depth and background.
		\section*{Declarations}
		\subsection*{Conflicts of interests/Competing interests}
		All authors declare that they have no conflicts of interest.
		\subsection*{Data availability statement}
		The data that support the findings of this study are available from the corresponding author upon reasonable request.
		
		\setcounter{equation}{0}
		\renewcommand{\theequation}{A\arabic{equation}}
		\section*{Appendix}
		\appendix
		\section{Stress components of various order of asymptotic series expansion}
		In order to assess the flow velocity, we need to first obtain the stress components from the Oldroyd-B constitutive equations after expanding the terms asymptotically. Here, for the inner layer, we have presented the normal and shear stress components for various order of corrections.\\
		
		\subsection{Constitutive relation for the $O(\beta^2)$ stress components}\label{sec:ASol_O(beta2)} 
		The constitutive relation for the $O(\beta^2)$ stress components in the inner layer becomes
		\begin{equation}
			\small
			\begin{aligned}
				\boxed{\widetilde \tau_{xx,02}}=0; \quad \boxed{\widetilde \tau_{xy,02}}=\frac{\partial \widetilde u_{02}}{\partial \widetilde y}; \quad \boxed{\widetilde \tau_{yy,02}}=2\frac{\partial \widetilde v_{02}}{\partial \widetilde y}
			\end{aligned}
		\end{equation}
		here, $\widetilde\tau_{xx,02}$ becomes zero and $\widetilde \tau_{xy,02}$ and $\widetilde \tau_{yy,02}$ have dependency on $O(\beta^2)$ velocity components only.
		\subsection{Constitutive relation for the $O(De\beta^2)$ stress components}\label{sec:ASol_O(Debeta2)} 
		We move towards obtaining the constitutive relation for the $O(De\beta^2)$ stress components in the inner layer, which reads
		\begin{equation}
			\small
			\begin{aligned}
				\boxed{\widetilde \tau_{xx,{12}}}=-4\frac{\partial \widetilde u_0}{\partial \widetilde x}\frac{\partial \widetilde u_{01}}{\partial \widetilde y}+2\frac{\partial \widetilde u_0}{\partial \widetilde x}\widetilde \tau_{xx,{02}}+2\frac{\partial \widetilde u_0}{\partial \widetilde y}\widetilde \tau_{xy,{02}}+2\frac{\partial \widetilde u_{01}}{\partial \widetilde x}\widetilde \tau_{xx,01}+2\frac{\partial \widetilde u_{01}}{\partial \widetilde y}\widetilde \tau_{xy,01}\\+2\frac{\partial \widetilde u_{02}}{\partial \widetilde x}\widetilde \tau_{xx,0}+2\frac{\partial \widetilde u_{02}}{\partial \widetilde y}\widetilde \tau_{xy,0}-\frac{\partial \widetilde \tau_{xx,0}}{\partial \widetilde x}\widetilde u_{02}-\frac{\partial \widetilde \tau_{xx,0}}{\partial \widetilde y}\widetilde v_{02}-\frac{\partial \widetilde \tau_{xx,01}}{\partial \widetilde x}\widetilde u_{01}-\frac{\partial \widetilde \tau_{xx,01}}{\partial \widetilde y}\widetilde v_{01}\\-\frac{\partial \widetilde \tau_{xx,{02}}}{\partial \widetilde x}\widetilde u_0-\frac{\partial \widetilde \tau_{xx,{02}}}{\partial \widetilde y}\widetilde v_0\\
				\boxed{\widetilde \tau_{xy,{12}}}=\frac{\partial \widetilde u_{12}}{\partial \widetilde y}-2\frac{\partial \widetilde u_0}{\partial \widetilde y}\frac{\partial \widetilde v_{01}}{\partial \widetilde y}+\frac{\partial^2 \widetilde u_0}{\partial \widetilde y \partial \widetilde x}\widetilde u_{01}+\frac{\partial^2 \widetilde u_0}{\partial \widetilde y^2}\widetilde v_{01}+\frac{\partial^2 \widetilde u_{01}}{\partial \widetilde y \partial \widetilde x}\widetilde u_{0}+\frac{\partial^2 \widetilde u_{01}}{\partial \widetilde y^2}\widetilde v_{0}\\+\frac{\partial \widetilde u_0}{\partial \widetilde y}\widetilde \tau_{yy,{02}}+\frac{\partial \widetilde u_{01}}{\partial \widetilde y}\widetilde \tau_{yy,01}+\frac{\partial \widetilde u_{02}}{\partial \widetilde y}\widetilde \tau_{yy,0}+\frac{\partial \widetilde v_0}{\partial \widetilde x}\widetilde \tau_{xx,{02}}+\frac{\partial \widetilde v_{01}}{\partial \widetilde x}\widetilde \tau_{xx,01}+\frac{\partial \widetilde v_{02}}{\partial \widetilde x}\widetilde \tau_{xx,0}\\-\frac{\partial \widetilde \tau_{xy,0}}{\partial \widetilde x}\widetilde u_{02}-\frac{\partial \widetilde \tau_{xy,0}}{\partial \widetilde y}\widetilde v_{02}-\frac{\partial \widetilde \tau_{xy,01}}{\partial \widetilde x}\widetilde u_{01}-\frac{\partial \widetilde \tau_{xy,01}}{\partial \widetilde y}\widetilde v_{01}-\frac{\partial \widetilde \tau_{xy,{02}}}{\partial \widetilde x}\widetilde u_0-\frac{\partial \widetilde \tau_{xy,{02}}}{\partial \widetilde y}\widetilde v_0\\
				\boxed{\widetilde \tau_{yy,{12}}}=2\frac{\partial \widetilde v_{12}}{\partial \widetilde y}+2\frac{\partial^2 \widetilde v_0}{\partial \widetilde y \partial \widetilde x}\widetilde u_{01}+2\frac{\partial^2 \widetilde u_0}{\partial \widetilde y^2}\widetilde v_{01}+2\frac{\partial^2 \widetilde v_{01}}{\partial \widetilde y \partial \widetilde x}\widetilde u_{0}+\frac{\partial^2 \widetilde v_{01}}{\partial \widetilde y^2}\widetilde v_0-8\frac{\partial \widetilde v_0}{\partial \widetilde y}\frac{\partial \widetilde v_{01}}{\partial \widetilde y}\\-2\frac{\partial \widetilde u_{01}}{\partial \widetilde y}\frac{\partial \widetilde v_0}{\partial \widetilde x}-2\frac{\partial \widetilde u_0}{\partial \widetilde y}\frac{\partial \widetilde v_{01}}{\partial \widetilde x}+2\frac{\partial \widetilde v_0}{\partial \widetilde x}\widetilde \tau_{xy,{02}}+2\frac{\partial \widetilde v_0}{\partial \widetilde y}\widetilde \tau_{yy,{02}}+2\frac{\partial \widetilde v_{01}}{\partial \widetilde x}\widetilde \tau_{xy,01}+2\frac{\partial \widetilde v_{01}}{\partial \widetilde y}\widetilde \tau_{yy,01}\\+2\frac{\partial \widetilde v_{02}}{\partial \widetilde x}\widetilde \tau_{xy,0}+2\frac{\partial \widetilde v_{02}}{\partial \widetilde y}\widetilde \tau_{yy,0}-\frac{\partial \widetilde \tau_{yy,0}}{\partial \widetilde x}\widetilde u_{02}-\frac{\partial \widetilde \tau_{yy,0}}{\partial \widetilde y}\widetilde v_{02}-\frac{\partial \widetilde \tau_{yy,01}}{\partial \widetilde x}\widetilde u_{01}-\frac{\partial \widetilde \tau_{yy,01}}{\partial \widetilde y}\widetilde v_{01}\\-\frac{\partial \widetilde \tau_{yy,{02}}}{\partial \widetilde x}\widetilde u_0-\frac{\partial \widetilde \tau_{yy,{02}}}{\partial \widetilde y}\widetilde v_0
			\end{aligned}
		\end{equation}
		here, the normal stress component $\widetilde \tau_{xx,{12}}$ is dependent on the leading order, $O(\beta)$, and $O(\beta^2)$ stress and velocity components. The other two stress components i.e. $\widetilde \tau_{xy,{12}}$ and $\widetilde \tau_{yy,{12}}$ depend on the $O(De\beta^2)$ velocity component along with the leading order, $O(\beta)$, and $O(\beta^2)$ stress and velocity components.
		\subsection{Constitutive relation for the $O(De^2)$ stress components}\label{sec:ASol_O(De2)}  
		We obtain the constitutive relation for the $O(De^2)$ stress components in the inner layer, which are
		\begin{equation}
			\small
			\begin{aligned}
				\boxed{\widetilde\tau_{xx,20}}=-\widetilde u_0\frac{\partial \widetilde \tau_{xx,10}}{\partial \widetilde x}-\widetilde u_{10}\frac{\partial \widetilde \tau_{xx,0}}{\partial \widetilde x}-\widetilde v_0\frac{\partial \widetilde \tau_{xx,10}}{\partial \widetilde y}-\widetilde v_{10}\frac{\partial \widetilde \tau_{xx,0}}{\partial \widetilde y}\\+2\widetilde \tau_{xx,0}\frac{\partial \widetilde u_{10}}{\partial \widetilde x}+2\widetilde \tau_{xx,10}\frac{\partial \widetilde u_{0}}{\partial \widetilde x}+2\widetilde \tau_{xy,0}\frac{\partial \widetilde u_{10}}{\partial \widetilde y}+2\widetilde \tau_{xy,10}\frac{\partial \widetilde u_{0}}{\partial \widetilde y}\\
				\boxed{\widetilde \tau_{xy,20}}=\frac{\partial \widetilde u_{20}}{\partial \widetilde y}-u_0\frac{\partial \widetilde \tau_{xy,10}}{\partial \widetilde x}-\widetilde u_{10}\frac{\partial \widetilde \tau_{xy,0}}{\partial \widetilde x}-\widetilde v_0\frac{\partial \widetilde \tau_{xy,10}}{\partial \widetilde y}-\widetilde v_{10}\frac{\partial \widetilde \tau_{xy,0}}{\partial \widetilde y}\\+\widetilde \tau_{xx,0}\frac{\partial \widetilde v_{10}}{\partial \widetilde x}+\widetilde \tau_{xx,10}\frac{\partial \widetilde v_{0}}{\partial \widetilde x}+\widetilde \tau_{yy,0}\frac{\partial \widetilde u_{10}}{\partial \widetilde y}+\widetilde \tau_{yy,10}\frac{\partial \widetilde u_{0}}{\partial \widetilde y}\\
				\boxed{\widetilde \tau_{yy,20}}=2\frac{\partial \widetilde v_{20}}{\partial \widetilde y}-\widetilde u_0\frac{\partial \widetilde \tau_{yy,10}}{\partial \widetilde x}-\widetilde u_{10}\frac{\partial \widetilde \tau_{yy,0}}{\partial \widetilde x}-\widetilde v_0\frac{\partial \widetilde \tau_{yy,10}}{\partial \widetilde y}-\widetilde v_{10}\frac{\partial \widetilde \tau_{yy,0}}{\partial \widetilde y}\\+2\widetilde \tau_{xy,0}\frac{\partial \widetilde v_{10}}{\partial \widetilde x}+2\widetilde \tau_{xy,10}\frac{\partial \widetilde v_{0}}{\partial \widetilde x}+2\widetilde \tau_{yy,0}\frac{\partial \widetilde v_{10}}{\partial \widetilde y}+2\widetilde \tau_{yy,10}\frac{\partial \widetilde v_{0}}{\partial \widetilde y}
			\end{aligned}
		\end{equation}
		here, the normal stress component $\widetilde\tau_{xx,20}$ is dependent on the leading order and $O(De)$ stress and velocity components. The other two stress components i.e. $\widetilde \tau_{xy,20}$ and $\widetilde \tau_{yy,20}$ depend on the $O(De^2)$ velocity component along with the leading order and $O(De)$ stress and velocity components.
		\subsection{Constitutive relation for the $O(De^2\beta)$ stress components}\label{sec:ASol_O(De2beta)}
		Then we proceed to obtain the constitutive relation for the $O(De^2\beta)$ stress components in the inner layer, which are of the form
		\begin{equation}
			\small
			\begin{aligned}
				\boxed{\widetilde \tau_{xx,{21}}}=-4\frac{\partial \widetilde u_0}{\partial \widetilde y}\frac{\partial \widetilde u_{10}}{\partial \widetilde y}+2\frac{\partial \widetilde u_0}{\partial \widetilde x}\widetilde \tau_{xx,1}+2\frac{\partial \widetilde u_0}{\partial \widetilde y}\widetilde \tau_{xy,1}+2\frac{\partial \widetilde u_{01}}{\partial \widetilde x}\widetilde \tau_{xx,10}+2\frac{\partial \widetilde u_{01}}{\partial \widetilde y}\widetilde \tau_{xy,10}\\+2\frac{\partial \widetilde u_1}{\partial x}\widetilde \tau_{xx,0}+2\frac{\partial \widetilde u_{1}}{\partial \widetilde y}\widetilde \tau_{xy,0}+2\frac{\partial \widetilde u_{10}}{\partial \widetilde x}\widetilde \tau_{xx,01}+2\frac{\partial \widetilde u_{10}}{\partial \widetilde y}\widetilde \tau_{xy,01}-\frac{\partial \widetilde \tau_{xx,0}}{\partial \widetilde x}\widetilde u_1-\frac{\partial \widetilde \tau_{xx,0}}{\partial \widetilde y}\widetilde v_1\\-\frac{\partial \widetilde \tau_{xx,01}}{\partial \widetilde x}\widetilde u_{10}-\frac{\partial \widetilde \tau_{xx,01}}{\partial \widetilde y}\widetilde v_{10}-\frac{\partial \widetilde \tau_{xx,1}}{\partial \widetilde x}\widetilde u_0-\frac{\partial \widetilde \tau_{xx,1}}{\partial \widetilde y}\widetilde v_{0}-\frac{\partial \widetilde \tau_{xx,10}}{\partial \widetilde x}\widetilde u_{01}-\frac{\partial \widetilde \tau_{xx,10}}{\partial \widetilde y}\widetilde v_{01}\\
				\boxed{\widetilde \tau_{xy,{21}}}=\frac{\partial \widetilde u_{21}}{\partial \widetilde y}-2\frac{\partial \widetilde u_0}{\partial \widetilde y}\frac{\partial \widetilde v_{10}}{\partial \widetilde y}-2\frac{\partial \widetilde u_{10}}{\partial \widetilde y}\frac{\partial \widetilde v_{0}}{\partial \widetilde y}+\frac{\partial^2 \widetilde u_0}{\partial \widetilde y \partial \widetilde x}\widetilde u_{10}+\frac{\partial^2 \widetilde u_0}{\partial \widetilde y^2}\widetilde v_{10}+\frac{\partial^2 \widetilde u_{10}}{\partial \widetilde y \partial \widetilde x}\widetilde u_{0}\\+\frac{\partial^2 \widetilde u_{10}}{\partial \widetilde y^2}\widetilde v_{0}+\frac{\partial \widetilde u_0}{\partial \widetilde y}\widetilde \tau_{yy,1}+\frac{\partial \widetilde u_{01}}{\partial \widetilde y}\widetilde \tau_{yy,10}+\frac{\partial \widetilde u_1}{\partial \widetilde y}\widetilde \tau_{yy,0}+\frac{\partial \widetilde u_{10}}{\partial \widetilde y}\widetilde \tau_{yy,01}+\frac{\partial \widetilde v_0}{\partial \widetilde x}\widetilde \tau_{xx,1}\\+\frac{\partial \widetilde v_{01}}{\partial \widetilde x}\widetilde \tau_{xx,10}+\frac{\partial \widetilde v_1}{\partial \widetilde x}\widetilde \tau_{xx,0}+\frac{\partial \widetilde v_{10}}{\partial \widetilde x}\widetilde \tau_{xx,01}-\frac{\partial \widetilde \tau_{xy,0}}{\partial \widetilde x}\widetilde u_1-\frac{\partial \widetilde \tau_{xy,0}}{\partial \widetilde y}\widetilde v_1-\frac{\partial \widetilde \tau_{xy,01}}{\partial \widetilde x}\widetilde u_{10}\\-\frac{\partial \widetilde \tau_{xy,01}}{\partial \widetilde y}\widetilde v_{10}-\frac{\partial \widetilde \tau_{xy,01}}{\partial \widetilde x}\widetilde u_0-\frac{\partial \widetilde \tau_{xy,1}}{\partial \widetilde y}\widetilde v_0-\frac{\partial \widetilde \tau_{xy,10}}{\partial \widetilde x}\widetilde u_{01}-\frac{\partial \widetilde \tau_{xy,10}}{\partial \widetilde y}\widetilde v_{01}\\
				\boxed{\widetilde \tau_{yy,{21}}}=2\frac{\partial \widetilde v_{21}}{\partial \widetilde y}+2\frac{\partial^2 \widetilde v_{10}}{\partial \widetilde y^2}\widetilde v_0+2\frac{\partial^2 \widetilde v_{0}}{\partial \widetilde y^2}\widetilde v_{10}-8\frac{\partial \widetilde v_0}{\partial \widetilde y}\frac{\partial \widetilde v_{10}}{\partial \widetilde y}-2\frac{\partial \widetilde u_{10}}{\partial \widetilde y}\frac{\partial \widetilde v_0}{\partial \widetilde x}-2\frac{\partial \widetilde u_0}{\partial \widetilde y}\frac{\partial \widetilde v_{10}}{\partial \widetilde x}\\-2\frac{\partial \widetilde u_0}{\partial \widetilde y}\frac{\partial \widetilde v_{01}}{\partial \widetilde x}-\frac{\partial \widetilde \tau_{yy,1}}{\partial \widetilde x}\widetilde u_0-\frac{\partial \widetilde \tau_{yy,1}}{\partial \widetilde y}\widetilde v_0-\frac{\partial \widetilde \tau_{yy,10}}{\partial \widetilde x}\widetilde u_{01}-\frac{\partial \widetilde \tau_{yy,10}}{\partial \widetilde y}\widetilde v_{01}-\frac{\partial \widetilde \tau_{yy,0}}{\partial \widetilde y}\widetilde v_1\\-\frac{\partial \widetilde \tau_{yy,{01}}}{\partial \widetilde x}\widetilde u_{10}-\frac{\partial \widetilde \tau_{yy,01}}{\partial \widetilde y}\widetilde v_{10}+2\frac{\partial \widetilde v_{01}}{\partial \widetilde y}\widetilde \tau_{yy,10}+2\frac{\partial \widetilde v_{1}}{\partial \widetilde y}\widetilde \tau_{yy,0}+2\frac{\partial \widetilde v_{10}}{\partial \widetilde x}\widetilde \tau_{xy,01}+2\frac{\partial \widetilde v_{10}}{\partial \widetilde y}\widetilde \tau_{yy,01}\\-\frac{\partial \widetilde \tau_{yy,0}}{\partial \widetilde x}\widetilde u_1+2\frac{\partial^2 \widetilde v_{0}}{\partial \widetilde y \partial \widetilde x}\widetilde u_{10}+2\frac{\partial^2 \widetilde v_{10}}{\partial \widetilde y \partial \widetilde x}\widetilde u_{0}+2\frac{\partial \widetilde v_0}{\partial \widetilde x}\widetilde \tau_{xy,1}+2\frac{\partial \widetilde v_0}{\partial \widetilde y}\widetilde \tau_{yy,1}+2\frac{\partial \widetilde v_{01}}{\partial \widetilde x}\widetilde \tau_{xy,10}
			\end{aligned}
		\end{equation}
		here, the normal stress component $\widetilde \tau_{xx,{21}}$ is dependent on the leading order, $O(De)$, $O(\beta)$ and $O(De\beta)$ stress and velocity components. The other two stress components i.e. $\widetilde \tau_{xy,{21}}$ and $\widetilde \tau_{yy,{21}}$ depend on the $O(De^2\beta)$ velocity component along with the leading order, $O(De)$, $O(\beta)$ and $O(De\beta)$ stress and velocity components.
		\subsection{Constitutive relation for the $O(De^2\beta^2)$ stress components}\label{sec:ASol_O(De2beta2)}  
		Then we proceed to obtain the constitutive relation for the $O(De^2\beta^2)$ stress components in the inner layer, which reads
		\begin{equation}
			\small
			\begin{aligned}
				\boxed{\widetilde \tau_{xx,{22}}}=-4\frac{\partial \widetilde u_0}{\partial \widetilde y}\frac{\partial \widetilde u_1}{\partial \widetilde y}-4\frac{\partial \widetilde u_{01}}{\partial \widetilde y}\frac{\partial \widetilde u_{10}}{\partial \widetilde y}+2\frac{\partial \widetilde u_{02}}{\partial \widetilde x}\widetilde \tau_{xx,10}+2\frac{\partial \widetilde u_0}{\partial \widetilde x}\widetilde \tau_{xx,{12}}+2\frac{\partial \widetilde u_{02}}{\partial \widetilde y}\widetilde \tau_{xy,10}\\+2\frac{\partial \widetilde u_{01}}{\partial \widetilde x}\widetilde \tau_{xx,1}-\frac{\partial \widetilde \tau_{xx,{12}}}{\partial \widetilde y}\widetilde v_0-\frac{\partial \widetilde \tau_{xx,{02}}}{\partial \widetilde x}\widetilde u_{10}+2\frac{\partial \widetilde u_{12}}{\partial \widetilde x}\widetilde \tau_{xx,0}+2\frac{\partial \widetilde u_{10}}{\partial \widetilde y}\widetilde \tau_{xy,{02}}+2\frac{\partial \widetilde u_{0}}{\partial \widetilde y}\widetilde \tau_{xy,{12}}\\-\frac{\partial \widetilde \tau_{xx,{12}}}{\partial \widetilde x}\widetilde u_0+2\frac{\partial \widetilde u_{12}}{\partial \widetilde y}\widetilde \tau_{xy,0}+2\frac{\partial \widetilde u_{10}}{\partial \widetilde x}\widetilde \tau_{xx,{02}}-\frac{\partial \widetilde \tau_{xx,01}}{\partial \widetilde y}\widetilde v_1-\frac{\partial \widetilde \tau_{xx,01}}{\partial \widetilde x}\widetilde u_1-\frac{\partial \widetilde \tau_{xx,1}}{\partial \widetilde x}\widetilde u_{01}\\+2\frac{\partial \widetilde u_1}{\partial \widetilde y}\widetilde \tau_{xy,01}+2\frac{\partial \widetilde u_{01}}{\partial \widetilde y}\widetilde \tau_{xy,1}-\frac{\partial \widetilde \tau_{xx,{02}}}{\partial \widetilde y}\widetilde v_{10}-\frac{\partial \widetilde \tau_{xx,10}}{\partial \widetilde x}\widetilde u_{{02}}-\frac{\partial \widetilde \tau_{xx,0}}{\partial \widetilde y}\widetilde v_{12}+2\frac{\partial \widetilde u_1}{\partial x}\widetilde \tau_{xx,01}\\-\frac{\partial \widetilde \tau_{xx,1}}{\partial \widetilde y}\widetilde v_{01}-\frac{\partial \widetilde \tau_{xx,10}}{\partial \widetilde y}\widetilde v_{{02}}-\frac{\partial \widetilde \tau_{xx,0}}{\partial \widetilde x}\widetilde u_{12}\\
				\boxed{\widetilde \tau_{xy,{22}}}=\frac{\partial \widetilde u_{22}}{\partial \widetilde y}+\frac{\partial^2 \widetilde u_{01}}{\partial \widetilde y \partial \widetilde x}\widetilde u_{10}+\frac{\partial^2 \widetilde u_{1}}{\partial \widetilde y \partial \widetilde x}\widetilde u_0+\frac{\partial^2 \widetilde u_{10}}{\partial \widetilde y \partial \widetilde x}\widetilde u_{01}+\frac{\partial \widetilde u_1}{\partial \widetilde y}\widetilde \tau_{yy,01}-\frac{\partial \widetilde \tau_{xy,1}}{\partial \widetilde x}\widetilde u_{01}\\-\frac{\partial \widetilde \tau_{xy,{02}}}{\partial \widetilde x}\widetilde u_{10}+\frac{\partial \widetilde v_1}{\partial \widetilde x}\widetilde \tau_{xx,01}-\frac{\partial \widetilde \tau_{xy,{12}}}{\partial \widetilde y}\widetilde v_{0}+\frac{\partial \widetilde u_{01}}{\partial \widetilde y}\widetilde \tau_{yy,1}+\frac{\partial \widetilde u_0}{\partial \widetilde y}\widetilde \tau_{yy,{12}}-\frac{\partial \widetilde \tau_{xy,0}}{\partial \widetilde x}\widetilde u_{{12}}\\+\frac{\partial \widetilde u_{02}}{\partial \widetilde y}\widetilde \tau_{yy,10}+\frac{\partial \widetilde u_{12}}{\partial \widetilde y}\widetilde \tau_{yy,0}-\frac{\partial \widetilde \tau_{xy,10}}{\partial \widetilde x}\widetilde u_{02}+\frac{\partial \widetilde u_{10}}{\partial \widetilde y}\widetilde \tau_{yy,{02}}-\frac{\partial \widetilde \tau_{xy,1}}{\partial \widetilde y}\widetilde v_{01}+\frac{\partial \widetilde v_{02}}{\partial \widetilde x}\widetilde \tau_{xx,10}\\+\frac{\partial \widetilde v_{01}}{\partial \widetilde x}\widetilde \tau_{xx,1}+\frac{\partial \widetilde v_{10}}{\partial \widetilde x}\widetilde \tau_{xx,{02}}+\frac{\partial \widetilde v_{12}}{\partial \widetilde x}\widetilde \tau_{xx,0}-\frac{\partial \widetilde \tau_{xy,01}}{\partial \widetilde x}\widetilde u_{1}-\frac{\partial \widetilde \tau_{xy,10}}{\partial \widetilde y}\widetilde v_{02}-\frac{\partial \widetilde \tau_{xy,{12}}}{\partial \widetilde x}\widetilde u_{0}\\+\frac{\partial^2 \widetilde u_1}{\partial \widetilde y^2}\widetilde v_{0}+\frac{\partial^2 \widetilde u_{10}}{\partial \widetilde y^2}\widetilde v_{01}+\frac{\partial^2 \widetilde u_0}{\partial \widetilde y^2}\widetilde v_{1}+\frac{\partial^2 \widetilde u_{01}}{\partial \widetilde y^2}\widetilde v_{10}-\frac{\partial \widetilde \tau_{xy,01}}{\partial \widetilde y}\widetilde v_1-\frac{\partial \widetilde \tau_{xy,0}}{\partial \widetilde y}\widetilde v_{12}\\-\frac{\partial \widetilde \tau_{xy,{02}}}{\partial \widetilde y}\widetilde v_{10}+\frac{\partial \widetilde v_0}{\partial \widetilde x}\widetilde \tau_{xx,{12}}-2\frac{\partial \widetilde u_{01}}{\partial \widetilde y}\frac{\partial \widetilde v_{01}}{\partial \widetilde y}
			\end{aligned}
		\end{equation}
		here,the stress components i.e. $\widetilde \tau_{xy,{22}}$ and $\widetilde \tau_{yy,{22}}$ are depend on the $O(De\beta^2)$ velocity component along with the leading order, $O(De)$, $O(\beta)$, $O(De\beta)$, $O(\beta^2)$, and $O(De\beta^2)$ stress and velocity components.

		\vspace{1cm}
		\hrule
		\hrule
		%\bibliography{Ref_higherorder}
		%\bibliographystyle{unsrtnat}% common bib file
		

%
\end{document}